\documentclass{PoS}

\usepackage{amsmath}

\title{Non-perturbative renormalization of four-quark operators and $B_K$ 
       with Schr\"{o}dinger functional scheme in quenched domain-wall QCD}

\ShortTitle{Non-perturbative renormalization of four-quark operators and $B_K$}

\author{\speaker{Y.~Nakamura}$^{a}$\thanks{E-mail: nakayou@het.ph.tsukuba.ac.jp}
         and Y.~Taniguchi$^{a,b}$ for CP-PACS Collaboration\\
        \llap{$^a$}Graduate School of Pure and Applied Sciences, University of Tsukuba, Tsukuba, Ibaraki 305-8571\\
        \llap{$^b$}Center for Computational Sciences, University of Tsukuba, Tsukuba, Ibaraki 305-8577\\
        }

\abstract{
We present non-perturbative renormalization factors for $\Delta S=2$ 
four-quark operators in quenched domain-wall QCD using the
Schr\"{o}dinger functional method.
Non-perturbative renormalization factor for $B_K$ is evaluated at
hadronic scale.
Combined with the non-perturbative RG running obtained by the Alpha
collaboration, our result yields renormalization factor which
converts lattice bare $B_K$ to the renormalization group invariant
one.
We apply the renormalization factor to bare $B_K$ previously obtained
by the CP-PACS collaboration with the quenched domain-wall QCD(DWQCD).
We compare our result with previous ones  obtained by perturbative renormalization
factors, different renormalization schemes or different quark actions. 
We also show that  chiral symmetry breaking effects in the
renormalization factor are numerically small.
}

\FullConference{The XXV International Symposium on Lattice Field Theory\\
		 July 30 - August 4 2007\\
		 Regensburg, Germany}

\begin{document}

\section{Introduction}

The Kaon $B$ parameter
\begin{equation}
B_K = \frac{\langle\overline{K}^0|
\bar{s}\gamma_\mu(1-\gamma_5)d\cdot\bar{s}\gamma_\mu(1-\gamma_5)d
|K^0\rangle}
{(8/3)\langle\overline{K}^0|
\bar{s}\gamma_\mu\gamma_5d|0\rangle\langle0|\bar{s}\gamma_\mu\gamma_5d
|K^0\rangle}
\end{equation}
is one of the fundamental weak matrix elements which have to be
determined theoretically for deducing $CP$ violation phase of the
Cabibbo-Kobayashi-Maskawa matrix from experiments.
Lattice QCD calculation may be an ideal tool to determine the
matrix element precisely from the first principle.
An essential step towards precise determination of $B_K$ is to
control systematic error in the renormalization factor.
Recently non-perturbative renormalization factor is preferably employed
to remove errors in the perturbative one.
Among several non-perturbative schemes on the lattice
the Schr\"odinger functional (SF) scheme \cite{SF} has an advantage that
systematic errors can be evaluated in a controlled manner.

A few years ago the CP-PACS collaboration calculated $B_K$ using
quenched domain-wall QCD (DWQCD) with Iwasaki gauge action
\cite{AliKhan:2001wr}.
Their result was renormalized perturbatively at one loop and have shown
a good scaling behavior with small statistical errors.
A main purpose of this paper is to  derive a non-perturbative renormalization factor
$\mathcal{Z}_{B_K}$ which convert the bare $B_K$ of the CP-PACS
collaboration to the renormalization group invariant (RGI) $\hat{B}_K$.
We adopt the SF scheme as an intermediate scheme to avoid systematic
uncertainties due to the finite lattice spacing.
The renormalization factor $\mathcal{Z}_{B_K}(g_0)$ is given at a fixed
bare coupling and its non-perturbative evaluation is decomposed into
three steps in the SF scheme as
\begin{equation}
\mathcal{Z}_{B_K}(g_0)=
Z_{VA+AV}^{\rm PT}(\infty,\mu_{\rm max})
Z_{VA+AV}^{\rm NP}(\mu_{\rm max},\mu_{\rm min})
Z_{B_K}^{\rm NP}(g_0,\mu_{\rm min}).
\end{equation}

We start from a renormalization factor at a low energy hadronic scale
$\mu_{\rm min}$
\begin{equation}
Z_{B_K}^{\rm NP}(g_0,\mu_{\rm min}) =
\frac{Z_{VV+AA}(g_0,\mu_{\rm min})}{Z_A^2(g_0)},
\end{equation}
where $Z_{VV+AA}$ is a renormalization factor for the parity even part
of the left-left four-quark operator and $Z_A$ is that for the axial
vector current.
A reason why we define the renormalization factor at  the hadronic scale is to suppress lattice
artifacts by a condition $a\mu_{\rm min}\ll1$.
This factor depends on both renormalization scheme and lattice
regularization.
Multiplying it with a lattice bare operator, the regularization dependence
is canceled and only the scheme dependence remains.

$Z_{VA+AV}^{\rm NP}(\mu_{\rm max},\mu_{\rm mix})$ represents
non-perturbative RG running of the parity odd part of the left-left four-quark
operators 
from the low energy scale $\mu_{\rm min}$ to
a high energy scale $\mu_{\rm max}=2^7\mu_{\rm min}$, where perturbation
theory can be safely applied. 
Among three steps
this requires the most extensive calculation.
Since this factor evaluated in the continuum limit does not depend on a specific lattice
regularization,
we can employ 
$Z_{VA+AV}^{\rm NP}(\mu_{\rm max},\mu_{\rm min})$ evaluated previously by
the Alpha collaboration with the improved
Wilson fermion action\cite{Guagnelli:2005zc}, 
instead of calculating it by ourselves. 
Even though
renormalization factors for the parity even and parity odd parts differ on the lattice
without chiral symmetry,
the difference disappears in the continuum limit, where the chiral symmetry is recovered. 

The last factor $Z_{VA+AV}^{\rm PT}(\infty,\mu_{\rm max})$ is the RG
evolution from the high energy scale $\mu_{\rm max}$ to infinity, which
absorb the scale dependence to give the RGI operator.
Since we are already deep in a perturbative region at $\mu_{\rm max}$ we
can evaluate this factor perturbatively.
Two loop calculation is given in Ref.~\cite{Palombi:2005zd}.
Note that scheme dependence is also canceled at this stage and the RGI
operator becomes scheme independent.

Our target in this study is the calculation of the first factor
$Z_{B_K}^{\rm NP}(g_0,\mu_{\rm min})$.
In order to further reduce the computational cost we use a
relation that $Z_{V}=Z_A$ implied by the chiral symmetry of DWQCD in SF scheme \cite{CPPACSZvZa03},
together with another one that  $Z_{VV+AA} = Z_{VA+AV}$, which will be numerically checked later,
and adopt
the following definition throughout this paper,
\begin{equation}
Z_{B_K}(g_0,\mu) = \frac{Z_{VA+AV}(g_0,\mu)}{Z_V^2(g_0)}.
\end{equation}

\section{Renormalization conditions for four-quark operator}
Since 
we utilize the step scaling function (SSF) obtained by the Alpha collaboration~\cite{Guagnelli:2005zc}
as an intermediate RG running factor from $\mu_{\rm min}$ to
$\mu_{\rm max}$,
the same renormalization scheme should be adopted for our calculation of 
$Z_{B_K}^{\rm NP}(g_0,\mu_{\rm min})$.
The renormalization condition is given by the following correlation
function~\cite{Guagnelli:2005zc}
\begin{equation}
\mathcal{F}_{\Gamma_A\Gamma_B\Gamma_C}(x_0) =
\frac{1}{L^3}\langle\mathcal{O}_{21}[\Gamma_A]\mathcal{O}_{45}[\Gamma_B]
O_{VA+AV}(x)\mathcal{O}'_{53}[\Gamma_C]\rangle,
\label{eqn:correlation}
\end{equation}
where subscripts $1\sim 5$ represent quark flavours and $O_{VA+AV}$ is the
parity odd four quark operator which consists of four different flavours from 1 to 4.
Boundary operators $\mathcal{O}_{ij}$ and $\mathcal{O}'_{ij}$ are given
in terms of boundary fields $\zeta$ and $\zeta'$ \cite{SF} as
\begin{equation}
\mathcal{O}_{ij}[\Gamma]=
a^6\sum_{\vec{x}\vec{y}}\bar{\zeta}_i(\vec{x})\Gamma\zeta_j(\vec{y}),
\hspace{0.5cm}
\mathcal{O}'_{ij}[\Gamma]=
a^6\sum_{\vec{x}\vec{y}}\bar{\zeta}'_i(\vec{x})\Gamma\zeta'_j(\vec{y}).
\label{eqn:boundary-operator}
\end{equation}
Due to the SF boundary condition for fermion fields the boundary operator
should be parity odd and we have two independent choices
$\Gamma=\gamma_5$ and $\Gamma=\gamma_k$ $(k=1,2,3)$.
For the correlation function to be totally parity-even we need at least
three boundary operators in (\ref{eqn:correlation}).
Logarithmic divergences in boundary fields $\zeta$'s can be removed
by the boundary-boundary correlation functions;
\begin{equation}
f_1=-\frac{1}{2L^6}
\langle\mathcal{O}'_{12}[\gamma_5]\mathcal{O}_{21}[\gamma_5]\rangle,
\quad
k_1=-\frac{1}{6L^6}\sum_{k=1}^{3}
\langle\mathcal{O}'_{12}[\gamma_k]\mathcal{O}_{21}[\gamma_k]\rangle .
\end{equation}
We adopt the following three choices
\cite{Guagnelli:2005zc}, whose perturbative expansions behave reasonably well
\cite{Palombi:2005zd}:
\begin{eqnarray}
h_1^\pm(x_0)=
 \frac{\mathcal{F}_{[\gamma_5,\gamma_5,\gamma_5]}(x_0)}{f_1^{3/2}},
\;
h_3^\pm(x_0)=
\frac{\frac{1}{3}\sum_{k=1}^3\mathcal{F}_{[\gamma_5,\gamma_k,\gamma_k]}(x_0)}
{f_1^{3/2}},
\;
h_7^\pm(x_0)=
\frac{\frac{1}{3}\sum_{k=1}^3\mathcal{F}_{[\gamma_5,\gamma_k,\gamma_k]}(x_0)}
{f_1^{1/2}k_1}.
\end{eqnarray}
We call three renormalization schemes defined through these correlation functions as
scheme 1, 3, 7 according to the Alpha collaboration
\cite{Guagnelli:2005zc}.

We impose the following renormalization condition
\begin{equation}
Z_{VA+AV;s}^\pm(g_0,\mu)h_s^{\pm}(x_0=L/2;g_0) =
 h_s^{\pm ({\rm tree})}(x_0=L/2)
\label{eqn:renormalization-condition}
\end{equation}
where $s$ labels the scheme. This means
that the renormalized correlation function should coincides with that at
tree level in the continuum at the middle of the box $x_0=L/2$.
The  renormalization scale at the low energy  (hadronic scale) is
introduced by the maximum box size $1/\mu_{\rm min}=2L_{\rm max}$,
where $L_{\rm max}$ is defined through renormalized coupling
$\bar{g}^2(1/L_{\rm max})=3.480$ in the SF scheme.
This box size corresponds to $L_{\rm max}/r_0=0.749(18)$~\cite{Takeda04}
in the continuum limit, so that $\mu_{\rm min}=1/2L_{\rm max}\sim263$ MeV using the Sommer scale
$r_0=0.5$ fm.

\begin{table}[t]
\begin{center}
 \begin{tabular}{c|c|c|c|c|c|c|c}
  \hline
  $N_L$ & $6$ & $8$ & $10$ & $12$ & $14$ & $16$ & $18$ \\
  \hline
  $\beta$ & $2.4446$ & $2.6339$ & $2.7873$ & $2.9175$ & $3.0313$ & $3.1331$ & $3.2254$ \\
  \hline
  $\mathcal{Z}_{B_K;1}(g_0)$ & $1.22(2)$ & $1.32(2)$ & $1.35(2)$ & $1.39(2)$ & $1.40(2)$ & $1.41(3)$ & $1.42(3)$   \\
  \hline
 \end{tabular}
\caption{$\beta$ values which gives the same box size $2L_{\rm max}$
 for each lattice sizes $N_L$.}
 \label{tab:beta-latsize}
\end{center}
\end{table}

\section{Numerical simulation details}

The CP-PACS collaboration has calculated the lattice bare $B_K$
in quenched DWQCD with Iwasaki gauge action at the domain wall height
$M=1.8$ and the fifth dimensional length $N_5=16$ \cite{AliKhan:2001wr}.
In order to renormalize this $B_K$ we adopted the same lattice
formulation as in the above.
The SF formalism for  Iwasaki gauge action is given in Ref.~\cite{Takeda04}.
For a domain-wall quark we adopted the orbifolding construction \cite{Taniguchi:2006qw}
to realize the SF boundary condition.
We employ the same size for the temporal and the spatial directions $T=L$.
We take the mass independent scheme in massless limit where all the
physical quark masses are set to zero.

Three lattice spacings $\beta=2.6$, $2.9$
and $3.2$\footnote{The data at $\beta=3.2$ is new and not published in
~\cite{AliKhan:2001wr}}
are employed corresponding to $a^{-1}\sim 2$, $3$ and $4$ GeV in the
previous simulation \cite{AliKhan:2001wr}.
In order to cover these three $\beta$'s we take $7$ lattice sizes.
At each lattice size we  tune $\beta$ to satisfy
$aN_L=2L_{\rm max}=1.498r_0$ using 
the following fit formula \cite{Takeda04}
\begin{equation}
\ln\left(\frac{a}{r_0}\right) = -2.193 -1.344(\beta-3) + 0.191(\beta-3)^2, 
\end{equation}
which covers $2.456\leq\beta\leq3.53$.
Lattice sizes and corresponding $\beta$ values are listed in
Table \ref{tab:beta-latsize}.

Quenched gauge configurations are generated by
using the HMC algorithm.
First $2000$ trajectories were discarded for thermalization.
We calculate the correlation functions on each configuration separated
by $200$ trajectories.
We employed $500$ to $1000$ configurations in this paper.

\section{Non-perturbative renormalization of $B_K$}

In this section we evaluate the renormalization factor
$\mathcal{Z}_{B_K;s}(g_0)$, which convert the lattice bare
$B_K(g_0)$ in DWQCD to the RGI $\hat{B}_K$.
Combining our renormalization factor
$Z_{B_K;s}^{\rm NP}(g_0,\mu_{\rm min})$ with the RG running factor
$Z_{VA+AV}^{\rm PT}(\infty,\mu_{\rm max})
Z_{VA+AV}^{\rm NP}(\mu_{\rm max},\mu_{\rm min})$
given by the Alpha collaboration, we obtain the renormalization factor
$\mathcal{Z}_{B_K;s}(g_0)$ at each $\beta$.
A result is given in Table \ref{tab:beta-latsize} for scheme $1$.
In order to obtain the renormalization factors at $\beta=2.6$, $2.9$ and
$3.2$ we fit it in a polynomial form, 
$\mathcal{Z}_{B_K;s}(g_0) = a_s + b_s(\beta-3) +c_s(\beta-3)^2$.

\begin{figure}[t]
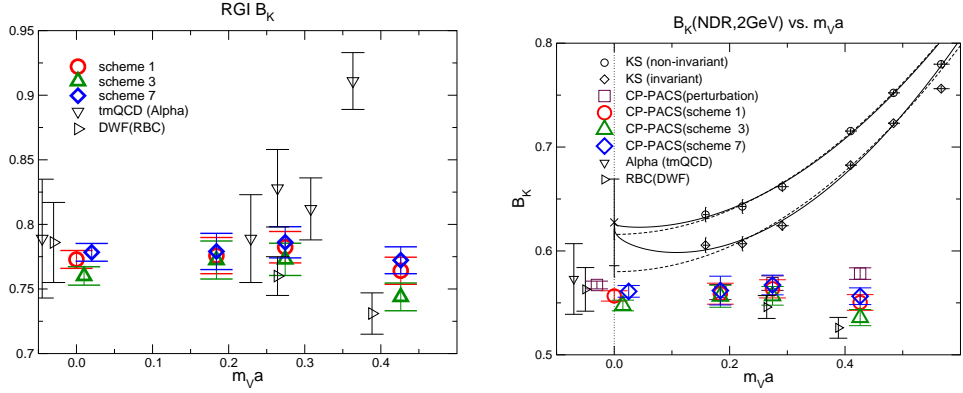

 \centering
 \includegraphics[width=60mm,clip]{fig/RGIBK.eps}
 \hspace*{5mm}
 \includegraphics[width=60mm,clip]{fig/BKNDR.eps}
 \caption{
Scaling behavior of RGI $\hat{B}_K$ (left panel) and
 $B_K(\overline{{\rm MS}},2{\rm GeV})$ (right panel).
 Our results are shown by open circle (scheme $1$), open up triangle
 (scheme $3$) and open diamond (scheme $7$).
 Previous results with tmQCD \cite{Guagnelli:2005zc} and DWF
 \cite{Aoki:2005ga} are also shown by open down and right triangles.
 In the right panel result of the KS fermion~\cite{Aoki:1997nr} and
 DWF~\cite{AliKhan:2001wr} with perturbative renormalizations are also
 given.
 The continuum results are slightly shifted to avoid overlaps.
} 
  \label{fig:result_BK}
\end{figure}

Multiplying it to the bare $B_K(g_0)$ we obtain the RGI $\hat{B}_K$,
whose scaling behavior is shown in the left panel of
Fig.~\ref{fig:result_BK}.
We also evaluate the renormalized $B_K$ in $\overline{\rm MS}$ scheme
with naive dimensional regularization (NDR) at a scale $\mu=2$ GeV.
The scaling behavior of $B_K^{\overline{\rm MS}}({\rm NDR},2{\rm GeV})$
is given in the right panel of Fig.~\ref{fig:result_BK}.
Since the scaling violation is negligible, it is reasonable to
take the continuum limit by a constant extrapolation.
We arrival at 
$\hat{B}_K = 0.773(7)(\begin{smallmatrix}+5 \\-13  \end{smallmatrix}) $
and
$B_K^{\overline{\rm MS}}({\rm NDR},2{\rm GeV}) =
0.557(5)(\begin{smallmatrix}+4 \\-10  \end{smallmatrix}) $,
where central values are taken from the scheme $1$,
the first parenthesis gives statistical error, and
upper(lower) error in the second parenthesis denotes difference between
schemes $1$ and $7$(1 and 3).
These are main results in this study.

\section{Additional investigations}

We perform two additional calculations.
One is to study the scaling behavior of SSF defined by
$\Sigma_{VA+AV;s}^{\pm}(u,a/L)=
\frac{Z_{VA+AV;s}^{\pm}(g_0,a/2L)}{Z_{VA+AV;s}^{\pm}(g_0,a/L)}
\bigl|_{\bar{g}^2(1/L)=u}$ in DWQCD.
The other  is to investigate a size of chiral symmetry breaking effects in
renormalization factors.

\subsection{Scaling behavior of SSF at $u=3.480$}

\begin{figure}[t]
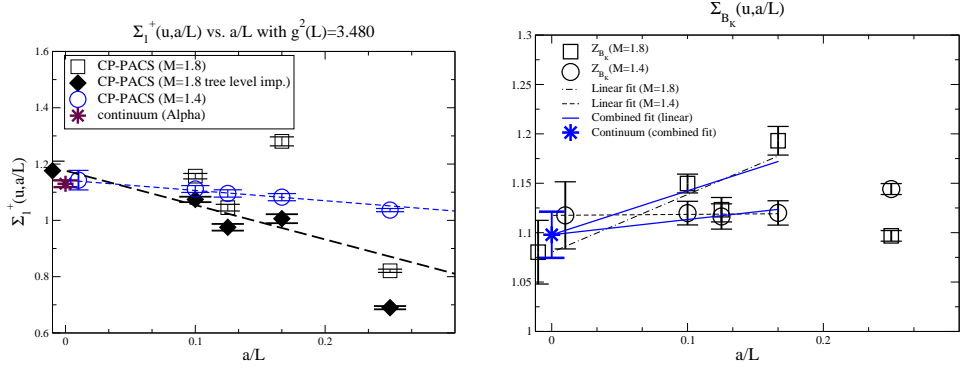

  \centering
  \includegraphics[width=60mm,clip]{fig/SSF.imp.PBC.eps}
  \hspace*{5mm}
  \includegraphics[width=60mm,clip]{fig/SSFzBK1.eps}
  \caption{Scaling behavior of SSF for $Z_{VA+AV;1}$(left panel) and
 for $Z_{B_K}$(right panel).
 Open squares are results for $M=1.8$ and open circles are for
 $M=1.4$ without improvement.
 Filled diamonds denote results for $M=1.8$ with tree level improvement.
 Star symbol denotes the continuum limit.}
  \label{fig:SSF}
\end{figure}

We investigate the SSF at strong coupling,
$u=\bar{g}^2(1/L_{\rm max})=3.480$, for four different $\beta$'s.
In the left panel of Fig.~\ref{fig:SSF} we show $a$
dependence of our SSF (open square) and the continuum limit of the Alpha
collaboration~\cite{Guagnelli:2005zc} (star).
We find that scaling violation of our SSF is large and oscillating at $M=1.8$.
We speculate that
this bad scaling behavior is caused by the $\mathcal{O}(a)$ boundary effect in
the SF scheme of  DWQCD, not by the $\mathcal{O}(a)$ bulk chiral symmetry breaking effect.
To see this we calculate the SSF at tree level,
where $N_5\to\infty$ limit is already taken.
We plot its scaling behavior by open circles in the left panel of
Fig~\ref{fig:tree.chiral}, where
the similar sort of oscillating behavior is observed at $M=1.8$. 
To exclude a possibility that the large scaling violation is caused by
the bulk chiral symmetry breaking effect,
we calculate $N_5$ dependence directly at $u=3.480$.
Indeed comparisons between $N_5=8$ and $N_5=16$ for 
$\Sigma_{VA+AV;s}(u,L/a=4)$ 
and between $N_5=32$ and $N_5=16$ for $\Sigma_{VA+AV;s}(u,L/a=6)$
show no $N_5$ dependence within statistical errors.

We  find that the scaling behavior of the tree SSF is much improved
at $M=0.9$, as shown 
in the left panel of Fig~\ref{fig:tree.chiral}. 
This suggests that the scaling behavior of the non-perturbative SSF is also
improved  when the renormalized domain-wall height is nearly equal to unity.
We first evaluate the SSF at $M=1.4$, where tadpole improved domain-wall
height is nearly equal to unity.
Result is given by open circles in the left panel of
Fig~\ref{fig:SSF}.
We find good scaling behavior at $M=1.4$,  and
the linear continuum extrapolation is 
consistent with the continuum limit of the Alpha collaboration.

For another improvement we introduce a tree level improvement, where
the continuum tree level correlation function $h_s^{\pm ({\rm tree})}(x_0)$ in the
renormalization condition (\ref{eqn:renormalization-condition}) 
is replaced to that on the lattice $h_s^{\pm ({\rm tree})}(x_0)^{\rm LAT}$.
We expect that the large scaling violation of the SSF is partly canceled by the tree SSF. 
We calculate the tree level improved SSF , where the tadpole improved value $M\sim 1.5$ 
of $M=1.8$ at each $\beta$ is used for $h_s^{\pm({\rm tree})}(x_0)$.
The result is given by the filled diamond in the left panel of
Fig~\ref{fig:SSF}.
We find that the scaling behavior is improved, so that we can perform a
linear extrapolation to the continuum limit, whose
value is consistent with that of the Alpha collaboration.

Furthermore we calculate the SSF of $B_K$, defined by 
$\Sigma_{B_K}(u,a/L)=\frac{Z_{B_K}(g_0,a/2L)}{Z_{B_K}(g_0,a/L)}$.
The result is plotted  in the right panel of Fig~\ref{fig:SSF} by
open square ($M=1.8$) and open circle $(M=1.4)$.
We find that a large scale violation in $\Sigma_1^+$ is partly canceled in $\Sigma_{B_K}$
at $M=1.8$.
Linear continuum extrapolations using data at finest three lattice spacings
give consistent values between $M=1.4$ and $M=1.8$.

\subsection{Chiral symmetry breaking effect}

\begin{figure}[t]
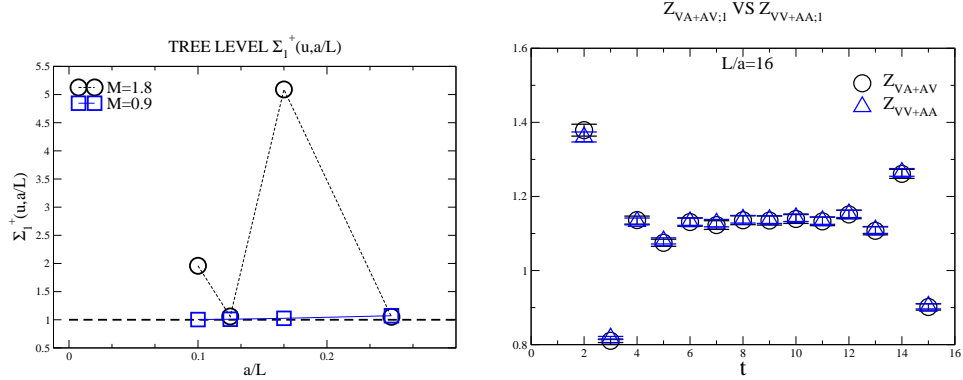

 \centering
 \includegraphics[width=60mm,clip]{fig/SSF.tree.eps}
 \hspace*{5mm}
 \includegraphics[width=60mm,clip]{fig/16z1pzvvaa.eps}
 \caption{Left panel is a scaling behavior of tree level SSF.
 Open circle shows results with $M=1.8$ and open square shows those with
 $M=0.9$.
 Right panel is a comparison between  $Z_{VA+AV}(g_0,\mu_{\rm min})$ 
 (open circle) and $Z_{VV+AA;1}(g_0,\mu_{\rm min})$ (open triangle) as
 a function of time $t$.
 We adopt scheme $1$ at the hadronic scale $\mu_{\rm min}$ on $L/a=16$
 lattice.}
  \label{fig:tree.chiral}
\end{figure}

We check whether  the chiral relation that $Z_{VV+AA}=Z_{VA+AV}$ we assumed 
is realized or not in our DWQCD.
If the chiral symmetry were exact we would have a chiral WT identity
\begin{eqnarray}
\langle{O}_{VA+AV}\mathcal{O}[\zeta]\rangle_{S} =
\langle{O}_{VV+AA}\tilde{\mathcal{O}}[\zeta]\rangle_{S}
\label{eqn:WTid}
\end{eqnarray}
under chiral rotation of the first flavour
$\delta q_1 = -i\gamma_5\tilde{q}$, $\delta\zeta_1=i\gamma_5\tilde{\zeta}_1$, 
$\delta\zeta'_1=i\gamma_5\tilde{\zeta}'_1$, where
$\tilde{\mathcal{O}}[\zeta]$ is a chiral rotation of some boundary
operator (\ref{eqn:boundary-operator}).
We get a relation $Z_{VV+AA} =Z_{VA+AV}$ from this WT identity.
On the other hand,  the domain-wall fermion action is not invariant under the
chiral rotation as
$S_{\rm dwf}\rightarrow S_{\rm dwf}+Y$, where $Y=\bar{\psi}X\psi$ is the
chiral symmetry breaking term at the  middle of the fifth dimension.
Therefore the WT identity becomes
$
\langle{O}_{VA+AV}\mathcal{O}[\zeta]\rangle_{S} =
\langle{O}_{VV+AA}\tilde{\mathcal{O}}[\zeta]\rangle_{S+Y} \neq
\langle{O}_{VV+AA}\tilde{\mathcal{O}}[\zeta]\rangle_{S},
$
and we estimate a possible chiral symmetry breaking effect by 
directly comparing
$\langle{O}_{VA+AV}\mathcal{O}[\zeta]\rangle_{S}$
with $\langle{O}_{VV+AA}\tilde{\mathcal{O}}[\zeta]\rangle_{S}$.
We evaluate the renormalization factor $Z_{VV+AA}(g_0,\mu_{\rm min})$
using the chirally rotated correlation function with 100 configurations
at each $\beta$.
The right panel of Fig~\ref{fig:tree.chiral} shows a time
dependence of $Z_{VA+AV}$ and $Z_{VV+AA}$ at $L/a=16$
We observe a good agreement between them, and a similar results are obtained at other $L/a$.
Therefore the relation $Z_{VV+AA}=Z_{VA+AV}$ is realized in our simulations.

This work is supported in part by Grants-in-Aid for Scientific Research
from the Ministry of Education, Culture, Sports, Science and
Technology(Nos.18740130).

\newcommand{\J}[4]{{#1} {\bf #2} (#3) #4}
\newcommand{\AP}{Ann.~Phys.}
\newcommand{\CMP}{Commun.~Math.~Phys.}
\newcommand{\EUR}{Eur.~Phys.J}
\newcommand{\IJMP}{Int.~J.~Mod.~Phys.}
\newcommand{\MPL}{Mod.~Phys.~Lett.}
\newcommand{\NP}{Nucl.~Phys.}
\newcommand{\NPSup}{Nucl.~Phys.~B (Proc.~Suppl.)}
\newcommand{\PL}{Phys.~Lett.}
\newcommand{\PR}{Phys.~Rev.}
\newcommand{\PRL}{Phys.~Rev.~Lett.}
\newcommand{\PTP}{Prog. Theor. Phys.}
\newcommand{\Suppl}{Prog. Theor. Phys. Suppl.}


\end{document}